\documentstyle[aaspp4,12pt,tighten,flushrt]{article}

\slugcomment{FERMILAB-Pub-96/067-A}

\input epsf.tex

\begin{document}
\def\la{\mathrel{\mathpalette\fun <}}
\def\ga{\mathrel{\mathpalette\fun >}}
\def\fun#1#2{\lower3.6pt\vbox{\baselineskip0pt\lineskip.9pt
        \ialign{$\mathsurround=0pt#1\hfill##\hfil$\crcr#2\crcr\sim\crcr}}}

\title{Weak Lensing and the Measurement of $q_0$ \\ from Type Ia Supernovae}

\author{Joshua A. Frieman}
\affil{NASA/Fermilab Astrophysics Center, Fermi National Accelerator 
Laboratory, \\ P.O.Box 500, Batavia, IL 60510}
\affil{Department of Astronomy 
and Astrophysics, University of Chicago, Chicago, IL 60637}

\begin{abstract} 
On-going projects to discover Type Ia supernovae 
at redshifts $z \sim 0.3 - 1$, coupled with improved techniques to  
narrow the dispersion in SN Ia peak magnitudes, have 
renewed the prospects for  
determining the cosmic deceleration parameter $q_0$. We estimate 
the expected uncertainty in the Hubble diagram 
determination of $q_0$ due to weak lensing by 
structure in the universe, which stochastically 
shifts the apparent brightness of distant 
standard candles. Although the 
results are sensitive to the density power spectrum 
on small scales, the induced flux dispersion $\sigma_m \la 0.04 
\Omega_{m,0}^{1/2}$ mag for sources at $z \leq 0.5$, well below the 
``intrinsic" spread of nearby SN Ia magnitudes, $\sigma_M \simeq 0.2$ mag.  
Thus, density inhomogeneities do not significantly impact
the current program to measure 
$q_0$, in contrast to a recent claim. If, however,  
light-curve shape and other calibrators can reduce the 
effective intrinsic spread to $0.1$ mag at high $z$, then weak lensing could  
increase the observed spread by 30 \%  in an $\Omega_{m,0}=1$ universe 
for SNe at $z \ga 1$.

\end{abstract}

\keywords{cosmology: large-scale structure of the universe, supernovae}

\section{Introduction} 
\label{sec:intro}

The determination of the cosmological parameters via the 
``classical" cosmological tests, such as the redshift-magnitude 
relation (Hubble diagram), remains a holy grail for observational 
astronomy. Past attempts to measure the deceleration parameter $q_0$ 
using galaxies as standard candles have foundered 
on the uncertainty in galaxy luminosity evolution (\cite{Tin72,Peeb93}).
Recently, there has been renewed hope that $q_0$ can be  
determined, thereby constraining the mean mass density 
and the cosmological constant, by using distant Type Ia supernovae 
(e.g., \cite{goob95}).  

Progress 
has come on two fronts. First, there is growing evidence that samples of 
SNe Ia, when suitably culled to excise peculiar lightcurves or spectra 
and cases of significant host-galaxy extinction, may provide 
reasonably good standard candles, with a dispersion $\sigma_M \simeq 0.3$ in 
peak absolute magnitude. Moreover, using the observed correlation between 
lightcurve shape and peak luminosity (\cite{ham95,rpk95}) as well as 
spectroscopic features that have been found to correlate with luminosity 
(\cite{bnf96,nugent95}),   
the effective dispersion of these `standardized candles' can apparently be 
reduced to $\sigma_M \simeq 0.1 - 0.2$ mag. 
Second, several groups have begun observing campaigns to discover a 
large number of high-redshift Ia supernovae, with coordinated follow-up 
programs to measure the lightcurves of SN candidates; 
more than 20 candidates at $z \sim 0.4 - 0.6$ have been found so far (e.g., 
\cite{perl95,schmidt95,perl96}). 

The case for SNe Ia as a  
population of `standard' candles  
is also on reasonable physical footing. While the details of the explosion 
mechanism remain poorly understood, there is general  
consensus that SNe Ia are thermonuclear explosions  
of accreting Carbon-Oxygen white dwarfs. 
On the other hand, it is not yet clear whether the white dwarf progenitors 
must be near the Chandrasekhar mass (e.g., \cite{hkwnt96}) or not 
(\cite{livarn95}), and thus the theoretical 
intrinsic spread in SN Ia luminosity is still a matter of some debate. 
In either case, there is optimism that, based on the correlations above, SNe Ia can be used 
to measure extragalactic distances and thereby to determine the 
cosmological parameters.

Recently, however, 
\cite{kvb95} have argued that large-scale structure provides 
a stumbling block to the accurate determination of $q_0$ from 
standard candles. A bundle of light rays from a distant 
source is sheared and focused due to 
deflection by intervening density inhomogeneities. 
Since gravitational lensing conserves surface 
brightness, the change in the cross-section of the bundle  
implies that the image of the source appears magnified 
(or demagnified) relative to a homogeneous universe. 
Using the ``swiss-cheese'' model of large-scale structure, Kantowski etal.  
find that the resulting change in received flux (apparent magnitude)  
could lead to a systematic underestimate of $q_0$ if one interprets the 
observations assuming a homogeneous universe. 
For example, for a source at $z=0.458$ 
(the redshift of SN 1992bi (\cite{perl95})), 
they find that the resulting error in $q_0$ can be as large as 
$\delta q_0 \simeq -0.33 q_0$. This claim is consonant 
with other studies which found that large-scale structure can 
significantly change the proportionality between  
angular-size distance and redshift, and therefore lead to 
difficulties, e.g., for the determination of 
$H_0$ from gravitational lens time delays  
(\cite{kant69,dyro72,aa85}, 1986, \cite{wata92,sasaki93}). 

In this Comment, we reevaluate this issue by estimating the rms 
fluctuation in the amplification of distant sources in a  
perturbed Friedmann-Robertson-Walker (FRW) universe. We find that 
the expected effects are smaller than those found by Kantowski, etal., 
of order several percent at most for 
sources at $z \la 0.5$, and subdominant in comparison to the $\sim 20$ \% 
intrinsic spread of nearby SN Ia magnitudes. 
Moreover, flux conservation implies that the magnification shift 
is random, with zero mean over all lines of 
sight (\cite{weinberg76}), not a systematic offset.
As a result, the 
amplification due to large-scale structure will not seriously  
impact the accuracy of $q_0$ measurements in current surveys. 

The primary reason for 
this different conclusion is that the swiss-cheese model
does not conform to our current understanding 
of the large-scale mass distribution of the universe. In particular, it 
does not accurately reflect the observational information gained    
from galaxy redshift and peculiar velocity surveys and the 
cosmic microwave background (CMB) anisotropy.  
This point has been made recently by \cite{seljak94} and Bar-kana (1995)
in the context 
of the determination of $H_0$ from QSO lens time delays. 
The argument that 
these effects should be small has also been made qualititatively by 
Peebles (1993). The issue  
was first laid out clearly by \cite{gunn67}, who showed that the 
rms fluctuation in the apparent brightness of a distant source can 
be expressed as a radial integral of the two-point correlation 
function of the mass distribution. Gunn's argument was updated 
to reflect the substantial advances in our understanding of 
large-scale structure in more recent numerical (\cite{jppg90}) 
and analytic (\cite{bl91}) studies (unbeknownst to the present 
author until this work was completed). In fact, Babul \& Lee's 
`DP' model is close to the model 
for the power spectrum that we adopt below, although there are 
some important quantitative differences which we outline later.

From the smallness of the CMB anisotropy on large scales 
to observations of galaxies and galaxy clusters on small scales, 
we have strong 
indications that the spacetime metric of the universe 
is well-described by a weakly perturbed FRW model. In the longitudinal 
gauge, the line element can be written

\begin{equation}
ds^2=a^2(\tau)\left[-(1+2\phi)d\tau^2+(1-2\phi)[d\chi^2+F^2(\chi)(d\theta^2
+\sin^2 \theta d\phi^2)]\right]
\label{metric}
\end{equation}

\noindent where $a(\tau)$ is the cosmic 
scale factor, $\chi$ denotes the comoving radial 
coordinate,  and $\tau=\int dt/a$ is 
the conformal time, with $\tau_0$ denoting the present. 
The function $F(\chi)$ depends on the 
spatial curvature $K$:   
$F(\chi)=K^{-1/2}\sin K^{1/2}\chi$ for $K>0$, $F=\chi$ 
for $K=0$, and $F=(-K)^{-1/2}\sinh(-K)^{1/2}\chi$ for $K<0$. The 
curvature can be expressed in terms of the present density parameter and 
the Hubble parameter $H_0=H(\tau_0)$,   
$K=(\Omega_0-1)H_0^2a^2_0$, where  
$\Omega = \Omega_m + \Omega_\Lambda$ 
includes both non-relativistic matter ($m$) 
and vacuum energy density (the cosmological constant $\Lambda$) and  
$\Omega_m = {\bar \rho}/
\rho_c = 8\pi G {\bar \rho}/3H^2$, with ${\bar \rho}(\tau)$ 
the mean density of matter. The metric perturbation variable 
$\phi$ is the relativistic analog of the Newtonian 
gravitational potential; over scales less than the Hubble
length $ H^{-1}$ it obeys the Poisson equation, 

\begin{equation}
	\nabla^2 \phi = {3\over 2} \Omega_m H^2 a^2 
	\delta
	\label{poisson}
\end{equation}

\noindent where the density contrast $\delta{({\bf x},\tau)} \equiv
\rho{({\bf x},\tau)}/ \bar \rho  - 1$. Observationally, the 
fluctuations in spatial geometry correspond to 
$\phi \la 10^{-4} - 10^{-5}$ 
from galaxy scales to the Hubble radius. 
We will assume $\phi \ll 1$ but do not place any restrictions on 
the amplitude of $\delta$. 

For a photon with direction $\hat{\bf n}$, the null geodesic equation in 
the metric (\ref{metric}) determines 
the rate of change in propagation direction 
due to inhomogeneities along the light path; to lowest order in $\phi$, 
the geodesic can be parametrized by the radial coordinate $\chi$, yielding 
$d{\hat{\bf n}}/d\chi = -2 {\bf \nabla}_\perp \phi$, where the derivative 
is transverse to the line of sight; we are implicitly assuming small 
deflection angles, so that the two-dimensional sphere can be approximated 
by a plane perpendicular to the unperturbed line of sight. Directions  in 
this plane can be parameterized by the two-dimensional angle $\mbox{\boldmath $\theta$}$.
A photon observed at 
direction $\mbox{\boldmath $\theta$}$ would have been seen at direction $\mbox{\boldmath 
$\beta$}=
\mbox{\boldmath $\theta$} + 
\mbox{\boldmath $\delta \theta$}$ in the absence of weak lensing effects, i.e., 
$\mbox{\boldmath $\delta \theta$}$ is the net transverse deflection of the light rays due 
to density fluctuations. From the geodesic equation, for a source 
at comoving distance $\chi_s$ we have 
(e.g., \cite{kaiser92,pynebirk94,villum95,seljak95}), defining 
$\mbox{\boldmath $\chi$}=-{\hat{\bf n}}\chi$, 

\begin{equation}
\mbox{\boldmath $\delta \theta$} = -2 \int_0^{\chi_s} {F(\chi_s-\chi) \over F(\chi_s)}
\mbox{\boldmath $\nabla$}_\perp \phi(\mbox{\boldmath $\chi$},\tau=\tau_0-\chi) d\chi ~ .
\label{deltheta}
\end{equation}

\noindent Here we have implicitly replaced the perturbed by the 
unperturbed path in the integral; while this is not a good 
approximation (e.g., \cite{FHS94}), we will only make use of the 
weaker 
assumption that the statistical properties of the potential field $\phi$ 
are identical along the perturbed and unperturbed 
paths (e.g., \cite{kaiser92}). 

The amplification of the observed 
image relative to the (unlensed) source is given by $A=1/{\rm det}~M$, 
where the $2\times 2$ amplification matrix   
$M_{ij} = \partial \beta_i/\partial \theta_j =
\delta_{ij} + \Phi_{ij}$, and   
$\Phi_{ij} =\partial \delta \theta_i/\partial \theta_j$. 
We can decompose the deformation tensor as   
$\Phi_{ij} = -\kappa\delta_{ij} + \gamma_{ij}$, 
where the trace ($\kappa$), the 
expansion, describes the uniform dilation or contraction of ray bundles, 
and the traceless part ($\gamma_{ij}$) describes their 
shear. In the limit of small deflection 
angles, we thus have 
$A=1/[(1-\kappa)^2-\gamma^2] \simeq 1+2\kappa$, where $\pm \gamma$ 
are the eigenvalues of $\gamma_{ij}$ and 
$\kappa = -(\Phi_{11}+\Phi_{22})/2$. The flux perturbation is 
therefore $\delta A = A-1 = -{\rm Tr}\Phi$. 
Using eqns. (\ref{poisson}, \ref{deltheta}), we find 

\begin{equation}
\delta A = {3(H_0 a_0)^2 \Omega_{m,0}\over F(\chi_s)} \int_0^{\chi_s} 
d\chi F(\chi) F(\chi_s - \chi) {a_0\over a(\tau_0-\chi)} 
\delta(\mbox{\boldmath $\chi$},\tau=\tau_0-\chi)
\end{equation}

\noindent where $\Omega_{m,0}$ is the present matter density parameter. 

To find the rms flux amplification along a random line of sight, 
we assume $\delta({\bf x},\tau)$ can be 
described as a continuous, homogeneous random process; this ignores  
discreteness in the mass density, an excellent approximation 
since the dark matter very 
likely consists of objects of mass less than $10^6 M_\odot$. 
It is convenient to Fourier-transform the density field, $\delta(\mbox{\boldmath $\chi$})=
(2\pi)^{-3}\int d^3k ~\delta({\bf k})\exp(i{\bf k}\cdot \mbox{\boldmath $\chi$})$, 
where 
the flat-space transform suffices for the small angles we 
are considering. Defining the density power spectrum via 
$\langle \delta({\bf k}) \delta^*({\bf k'})\rangle = (2\pi)^3 P({\bf k}) 
\delta^3_D({\bf k}-{\bf k'})$, we have (defining $\sigma^2_A = \langle (\delta A)^2\rangle$)

\begin{equation}
\sigma_A^2 = 9\pi \Omega_{m,0}^2 (H_0 a_0)^4 
 \int_0^{\chi_s} d\chi F^2(\chi)
{F^2(\chi_s - \chi)\over F^2(\chi_s)} (1+z(\tau))^2
\int {\Delta^2(k,z)\over k} {dk \over k}
\label{dA2}
\end{equation}

\noindent where $\Delta^2(k,z) = k^3P(k,z)/2\pi^2 = d\sigma^2_\delta/d\ln k$ is
the contribution per logarithmic wavenumber interval to the variance of 
the density field. Here, $z(\tau)$ is the redshift at epoch $\tau=\tau_0-\chi$, 
given implicitly by 

\begin{equation}
\chi(z)={1\over H_0 a_0} \int_{0}^{z} {dz'\over \left[\Omega_{m,0}(1+z')^3 
+(1-\Omega_{m,0}-\Omega_\Lambda)(1+z')^2 + \Omega_\Lambda \right]^{1/2}} ~.
\label{chiz}
\end{equation}

\noindent For $\Omega_\Lambda = 0$, this has a well-known analytic solution  
for $z(\chi)$ (\cite{mattig58}). 
In (\ref{dA2}), we have used a small-angle approximation, so that   
only waves nearly perpendicular to the line of sight 
contribute to the amplification (\cite{b91,kaiser92}); in addition, 
we have assumed that the density correlation length is small 
compared to the Hubble radius, so we only include contributions to 
$\langle \delta({\bf k},\tau) \delta^*({\bf k'},\tau')\rangle$ from 
equal times, $\tau=\tau'$. 
  
Over a limited range of wavenumber $k$ and redshift $z$, 
the power spectrum scales with 
time as $\Delta^2(k,z)=\Delta^2(k)(1+z)^{-\epsilon}$, separating 
the integrals in (\ref{dA2}).  
At small $k$, $\Delta \ll 1$, and it obeys linear perturbation 
theory, $\epsilon=2$. 
At large $k$, where $\Delta \gg 1$, 
the clustering is highly non-linear and 
has approximately reached virial equilibrium, i.e., 
galaxies and groups of galaxies reach a fixed physical size; in this 
`stable clustering' regime, if the two-point 
density correlation function obeys $\xi(r) \sim r^{-\gamma}$, then 
$\epsilon=3-\gamma$. Since the observed {\it galaxy} correlation function 
is a power-law with $\gamma \simeq 1.8$ at $r < 10h^{-1}$ Mpc, 
we expect $\epsilon \simeq 1.2$ on small scales, although 
with the caveat that the galaxy and mass distributions may differ 
significantly on these scales (see below). A third more radical possibility is 
that the observed clustering is merely ``painted on" and expands with the 
Hubble flow; in this case, $\epsilon=0$. For the numerical estimates 
below, we will thus consider the  
range $\epsilon= 0 - 2$. Since the rms amplification is dominated 
by structure in the non-linear regime, we expect $\epsilon \simeq 1.2$ 
to be the most accurate representation of the evolution. Recent 
N-body simulations confirm that the growth factor is intermediate 
between the stable clustering and linear regimes on small scales 
at recent epochs (\cite{ccc96}), while it may be faster than linear 
($\epsilon > 2$) on intermediate scales; in either case, the results 
presented below constitute an upper limit on the lensing effect.

To model the present power-spectrum on small scales, we could simply 
Fourier transform the 
galaxy correlation function, $\xi_g(r)=(r/r_0)^{-1.8}$, 
where the galaxy correlation 
length $r_{0,g} \simeq 5.4h^{-1}$ Mpc. However, 
this does not take into account 
the {\it bias} between the galaxy and mass distributions, i.e., 
the likelihood that light does not trace mass on these scales.
To remedy this, we can incorporate dynamical information.
From the (simplified version of the) 
cosmic virial theorem (Peebles 1980, 1993), 
the predicted pairwise velocity 
dispersion on small scales is roughly $\sigma^2_v(r) \sim [3/2(3-\gamma)]
H^2_0 \Omega_{m,0} r^2 (r_0/r)^\gamma$. Assuming the matter correlation 
function has the same slope as that of the galaxies, this yields 
an estimate for the density correlation length, $r_0 \simeq 
5.4h^{-1} ~{\rm Mpc} (0.1/\Omega_{m,0})^{0.55} (\sigma_v(1h^{-1}~{\rm Mpc})/
300 ~{\rm km/sec})^{1.1}$. The standard 
estimates of the galaxy pairwise velocity 
dispersion at $r \simeq 1 h^{-1}$ Mpc separation have been 
$\sigma_v \sim 300$ km/sec (\cite{dp83,bean83}). Recent redshift surveys, 
however, have yielded higher values, 
$\sigma_v \sim 500 - 700$ km/sec 
(\cite{guzzo95,marzke95}). Moreover, Somerville, Davis, \& Primack (1996) and 
Somerville, Primack, \& Nolthenius (1996) have found that estimates 
of $\sigma_v(r)$ in both simulations and galaxy catalogs show large 
scatter. However, when the cores of rich clusters, which contain only a  
small fraction of the mass, are excluded, it appears that $\sigma_v 
\sim 300$ km/sec. In any case, to obtain an upper bound for the weak 
lensing effect, we will take $\sigma_v(1 h^{-1} ~{\rm Mpc}) = 650$ 
km/sec, implying $\Delta^2(k) \simeq 8.7(k/h~{\rm Mpc}^{-1})^{1.8}
\Omega_{m,0}^{-1}(\sigma_v(1)/650)^2$.
Note that the assumption made here 
that the galaxy and density correlation functions 
have the same small-scale slope is not 
well motivated on highly non-linear scales, and the results below 
should be interpreted with this caveat.  

For this model of the power spectrum, and 
in general for $\gamma > 1$, the $k$-integral in (\ref{dA2}) diverges in the 
ultraviolet and must be cut-off. Physically, the 
slope of the density correlation function must fall below unity 
below some length scale (even though 
the {\it galaxy} correlation function is an unbroken $\gamma \simeq 1.8$ 
power law down to $r \sim 10 h^{-1}$  
kpc). This flattening is expected to happen at least at the scale of 
individual galaxy halos (\cite{kaiser92}): below this scale, $\xi(r)$ is  
dominated by correlations within individual halos, yielding 
$\xi(r) \sim r^{-(2\nu-3)}$ for halos with density profile $\rho(r) 
\sim r^{-\nu}$ (\cite{peebles74,mcsilk77,shjain96}); for nearly isothermal 
halos, $\nu \simeq 2$, the corresponding slope is $\gamma \simeq 1$. Up to 
logarithmic corrections, we model this effect 
by imposing a cutoff in (\ref{dA2}) at the approximate halo scale, 
$k_c = 1/(100 h^{-1}~{\rm kpc})$. Using this model in (\ref{dA2}), the 
rms flux perturbation at fixed 
$z_s$ scales approximately as $\Omega_{m,0}^{1/2}(\sigma_v(1)/650)(k_c/
10 h ~{\rm Mpc}^{-1})^{0.4}$.   

Fig.1 shows the dispersion in apparent magnitude for distant standard 
candles as a function of redshift, 
$\sigma_m = 1.086\sigma_A$ mag, for 
cosmological models with $\Omega_{m,0}=0.1$, 0.3, and 1. 
For $\Omega_{m,0}=1$, we show results for $\epsilon = 0$, 1.2, and 2 
to bracket the plausible range of evolution models. For the other  
cases, we show results for $\epsilon=1.2$ only;  
since structure formation freezes out early in $\Omega_{m,0}<1$ models (at 
$1+z_f \simeq \Omega_{m,0}^{-1}$), 
stable clustering should be a reliable prescription here. 
For $\Omega_{m,0}=0.3$, we show results for  
a spatially flat model with non-zero cosmological 
constant, $\Omega_\Lambda = 0.7$, and for an open model with $\Lambda=0$. 

Since the imposition of a sharp cutoff in the density power spectrum 
appears to be a rather crude approximation, as a check we have also explored 
models for the small-scale power spectrum based on    
the cold dark matter (CDM) model and 
CDM with non-zero $\Lambda$, with scale-invariant primordial 
fluctuations, extended into the 
non-linear regime using scaling formulae derived from N-body 
simulations (\cite{ham91,pd94,jmw95,bg96,pd96}). In these models, 
$P(k)$ scales as $k$ at small wavenumber but turns over to 
$k^{-3}$ at large $k$, yielding much better convergence 
for the flux dispersion $\sigma_A$. We normalize such models by 
requiring that they reproduce 
the observed 
galaxy cluster X-ray temperature distribution function according to 
the predictions of 
Press-Schechter theory (\cite{wef93}); for CDM models, 
this corresponds approximately to imposing the 
constraint that the linear theory  
rms mass fluctuation in a sphere of radius 8 $h^{-1}$ Mpc is 
$\sigma_8 = 0.6 \Omega_{m,0}^{-C(\Omega_{m,0},\Lambda)}$, where 
$C=0.36+0.31\Omega_{m,0}-0.28\Omega^2_{m,0}$ for open models ($\Lambda=0$) 
and $C=0.59-0.16\Omega_{m,0}+0.06\Omega^2_{m,0}$ for spatially 
flat (non-zero $\Lambda$) models (\cite{vl}). 
The results for the amplification dispersion in these cluster-normalized 
CDM models 
are shown in Fig. 2 for the same model parameters as in Fig. 1. The 
results for $\sigma_A$ agree reasonably well with those of   
the power-law model, although they are somewhat higher for the models 
with $\Omega_{m,0} <1$. This difference is due in part to the fact 
that the cluster normalization yields more small-scale power in 
these models than the cosmic virial theorem normalization for the 
corresponding power-law model. Thus, while 
the uncertainties in $\sigma_v$ and in the accuracy of the cosmic 
virial theorem are still significant, this comparison suggests 
that our estimate for $\sigma_A$ should be accurate to within a factor 
of two.  

The implications of weak lensing for the determination of $q_0$
follow directly from Figs. 1 and 2. For sources at $z \leq 0.7$, 
$\sigma_m \leq 0.06$, well below 
the ``intrinsic'' $0.2 - 0.3$ mag spread in nearby SN Ia magnitudes. 
At $z \ll 1$, from the redshift-magnitude relation, 
the resulting `$1\sigma$' uncertainty in the deceleration parameter, 
$q_0=\Omega_{m,0}/2-\Omega_\Lambda$, for a single 
source at redshift $z$ is $\sigma_{q_0} \simeq \sigma_A/z$.
For example, for $z=0.5$ and $\Lambda = 0$, we find $\sigma_{q_0} \simeq 
0.1 q_0^{1/2}$  
($\la 0.07$ for $\Omega_{m,0} \leq 1$).  In the future, if SN Ia searches 
discover sources at $z \ga 1$, then weak lensing may be a significant factor. 
In particular, if light-curve shape and other correlators 
with peak magnitude can reduce the effective intrinsic 
spread to 0.1 mag (even at high $z$), then 
density fluctuations could increase 
the observed dispersion in an $\Omega_{m,0} =1$ universe 
for sources at $z = 1$  
by of order 30\%. Since the amplification is caused by 
the foreground mass distribution, one could in principle use the 
angular correlation function of $\delta A$ to probe the large-scale 
mass power spectrum; in practice, this will require thousands of 
well-measured SNe Ia at redshifts $z \ga 0.5$ spread over hundreds 
of square degrees. This may be possible with the Next Generation Space 
Telescope.

The results shown here can be compared to those of Babul \& Lee (1991). 
While they considered 
only the Einstein-de Sitter $\Omega_{m,0}=1$ case, we have presented  
results for arbitrary $\Omega_m$ and $\Lambda$. 
Quantitatively, 
their `DP' model for the power spectrum is quite similar to that 
used here, but it overestimated the rms amplification $\sigma_A$  
because it did not include the constraint from galaxy pairwise velocities on 
small scales. They also presented results for the $\Omega=1$ CDM model 
with bias factor $b \simeq 2.5$; this normalization, which reflected earlier,  
lower estimates of the small-scale pairwise velocity dispersion, is now 
known to be 
unacceptably low and therefore substantially underestimates $\sigma_A$. 
Thus, while their results bracket those here, the estimates in Figs. 1 
and 2 should be more accurate.  

A final issue of concern for these  surveys is amplification bias.   
A fraction of the SNe in a magnitude-limited 
survey are strongly amplified ($\delta A \gg \sigma_A$) by foreground mass 
concentrations very close to the line of sight; this  
would cause large systematic errors in the estimate of $q_0$. However, 
the lensing galaxy or cluster responsible could generally be seen and 
such events removed from the sample. Even if the lenses are too faint 
to be detected, the probability for such strong lensing events at 
moderate redshift, $z_s \la 1$, is known to be very small, 
based on the low incidence of multiply imaged QSOs. While the optical  
depth $\tau$ for significant amplification by foreground galaxies 
(e.g., $\delta A \ga 0.1$) is much higher 
than that for multiple imaging, it is still quite low. For example, for 
$\Omega_{m,0}=1$ and assuming isothermal 
galaxy halos extend to $r \ga 50 h^{-1}$ kpc, integration over the galaxy 
luminosity function 
with the Faber-Jackson relation gives $\tau(\delta A > 0.1, z_s=1) 
\simeq 6.5\times 10^{-3}$ 
for amplification by individual foreground galaxies; for a non-zero 
cosmological constant satisfying $\Omega_\Lambda = 1 - \Omega_{m} < 0.8$, 
$\tau(>0.1,1) < 0.02$. 
The result of these near encounters 
is a very small non-Gaussian tail in the amplification distribution at 
large $\delta A$. On the other hand, if a substantial fraction of 
the cosmic density is in compact objects, the high amplification 
tail can be much more significant (\cite{sw87,lsw88,rauch91}).  
If high amplification ($\ga 1$ mag) events are not soon found 
in the high-redshift SN Ia searches, one will be able to constrain 
the contribution to $\Omega$ from objects of mass $\ga 10^{-3}M_\odot$. 

For completeness, I note that weak lensing does not affect the shape 
of SN Ia lightcurves: for a source at redshift $z_s$, 
the fall-off from the peak is given by the usual time dilation factor,  
$\Delta t= \Delta t_i (1+z_s)$, where $\Delta t_i$ is the fall-off 
timescale in the SN rest frame. 

{\it Note added:} After this work was accepted for publication, two 
recent numerical efforts have clarified the weak lensing effects of 
large-scale structure (\cite{wamost}, D. E. Holz and R. M. Wald, 
to be published). Using ray-shooting 
techniques through N-body and phenomenologically constructed universes, 
it has been directly confirmed that the mean flux from a distant standard 
candle is that given by the standard FRW luminosity distance. However, 
the distribution of the amplification is not symmetric about this mean. 
The majority of lines of sight pass through regions underdense compared 
to the mean, leading to modest de-amplification, while a smaller number 
of rays pass near dense mass concentrations and suffer substantial 
amplification. Thus, for a survey with a finite number of sources, 
there can be a small bias in the results for $q_0$, but the effect 
is very small at the current redshifts ($z \la 0.6$) probed by the 
SN Ia surveys.


\acknowledgments

I thank D. Holz, A. Olinto, S. Perlmutter, U. Seljak, J. Silk, A. Stebbins, 
and R. Wald 
for conversations, and the Institute for 
Nuclear and Particle Astrophysics (LBL) and the Center for Particle Astrophysics 
(UC Berkeley) for hospitality while this work was being completed. 
This research was supported in part 
by the DOE and by NASA grant NAG5-2788 at Fermilab.


\clearpage

\begin{figure}[t!]
\centering
\centerline{\epsfxsize=17. truecm \epsfysize=17. truecm 
\epsfbox{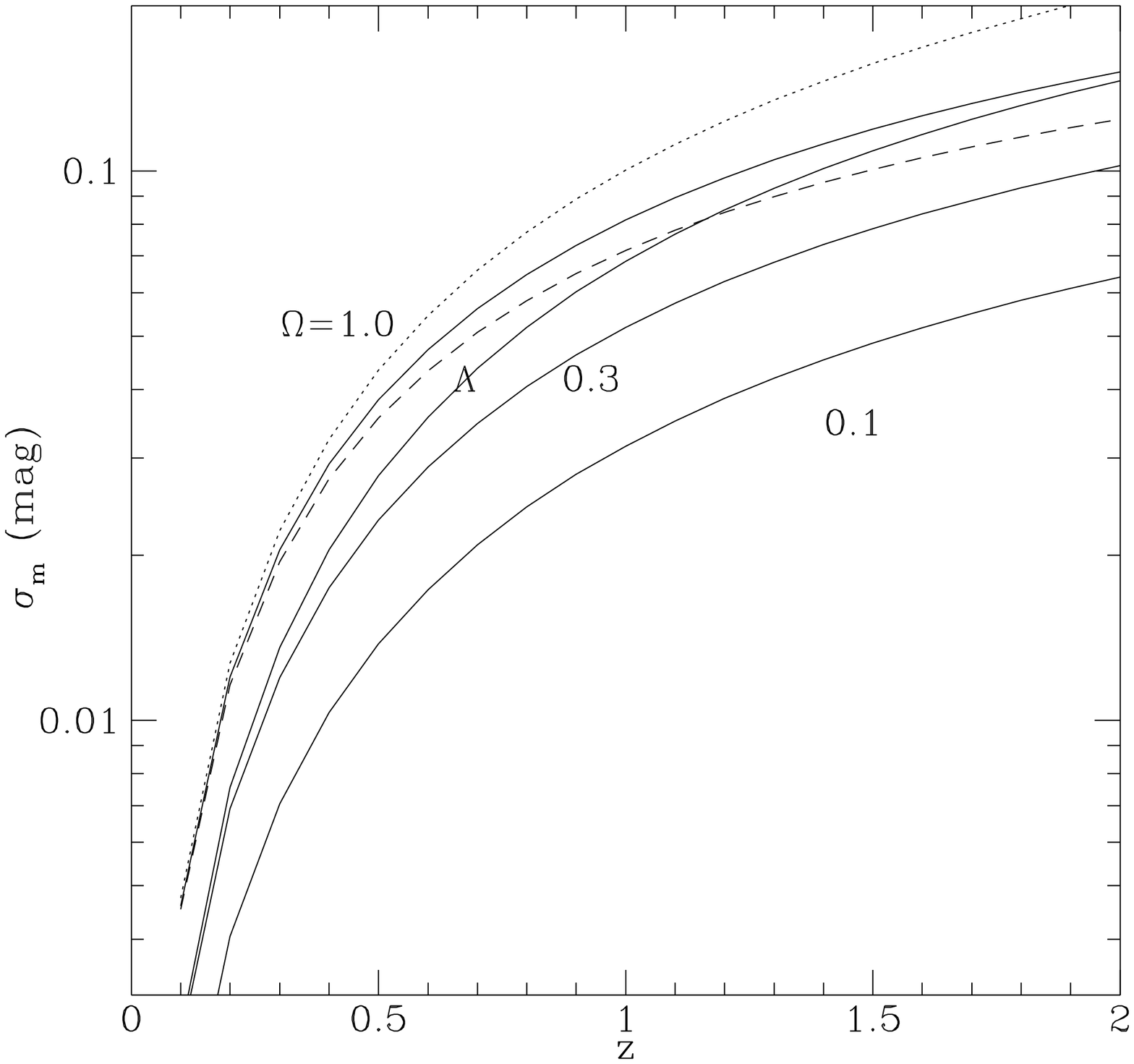}
}

\vspace{1.5 truecm}

\caption{Fig. 1. Dispersion in flux (in magnitudes) for a 
source at redshift $z$, for $\Omega_{m,0} =0.1$, 0.3, 1.0.
Solid curves assume stable clustering, $\epsilon=1.2$. 
For $\Omega_{m,0}=1$, dashed curve corresponds 
to linear theory evolution, $\epsilon=2$, and 
dotted curve corresponds to `painted on' structure ($\epsilon=0$).
Lower curves marked $\Omega_{m,0}=0.1, 0.3$ are open models; curve 
marked `$\Lambda$' is a flat universe 
with $\Omega_\Lambda=0.7$. The small-scale power spectrum has been cut off at 
$k_c=1/(100 h^{-1} 
~{\rm kpc})$; the dispersion scales as $\sigma \propto k_c^{0.4}$.}
\label{p1lnm1}
\end{figure}

\clearpage

\begin{figure}[t!]
\centering
\centerline{\epsfxsize=17. truecm \epsfysize=17. truecm 
\epsfbox{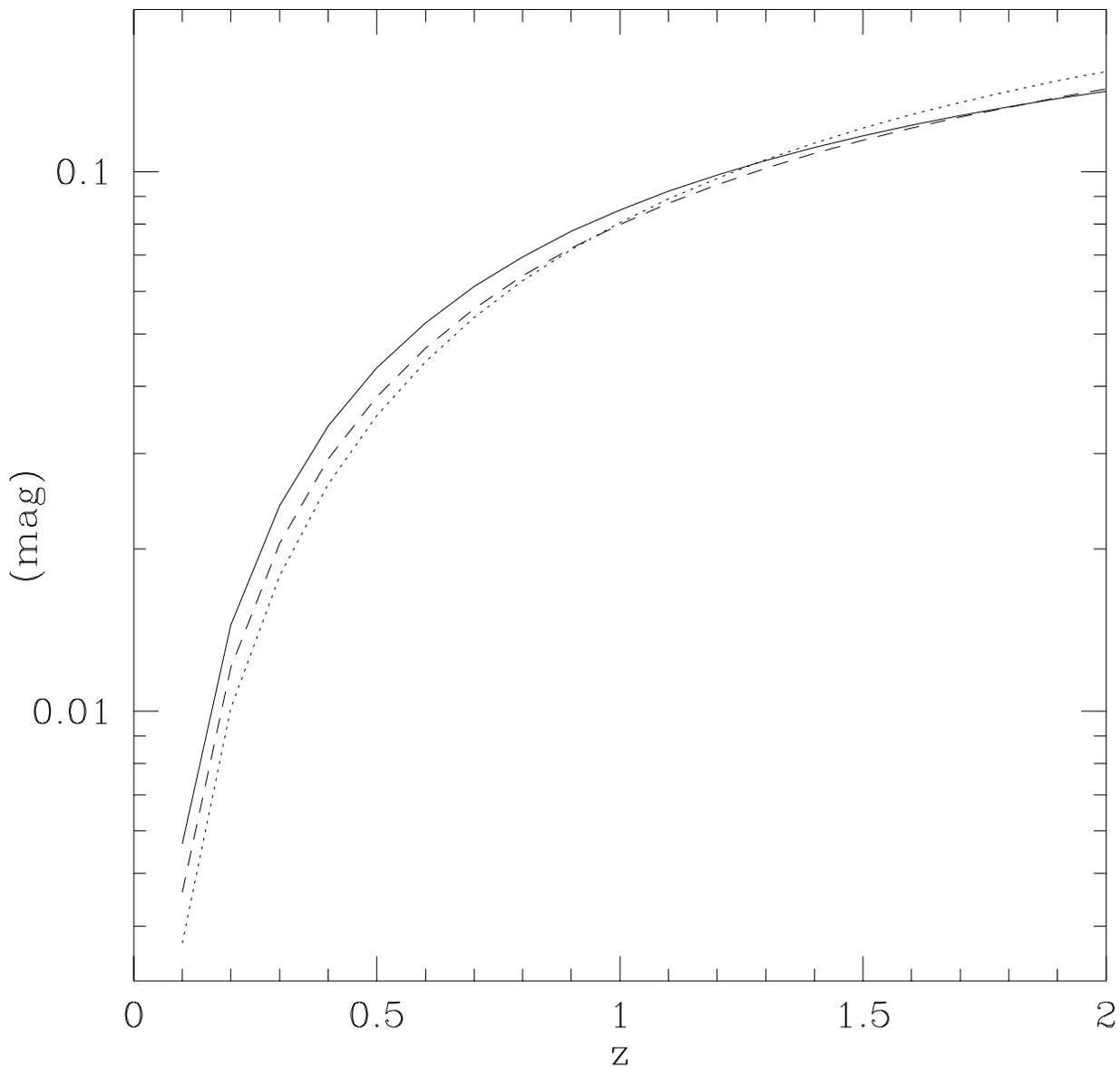}
}

\vspace{1.5 truecm}

\caption{Fig. 2. Flux dispersion vs. redshift for models with the 
same cosmological parameters as in Fig. 1, but for the CDM model 
of structure formation instead of the phenomenological power-law model. 
The models are standard CDM $\Omega_{m,0}=1$, $h=0.5$ (solid), 
a flat $\Lambda$CDM model with $\Omega_\Lambda=0.7$, $h=0.7$ (dotted), 
and an open CDM model with $\Omega_{m,0}=0.3$, $h=0.7$ (dashed).}

\label{p2}
\end{figure}

\end{document}